# ARTICLE

# Ionization energy and electron affinity of fullerene $C_{60}$ in the Hubbard model in the static fluctuation approximation.

Gennadiy Ivanovich Mironov*[a]

Within the Hubbard model, the ionization energy and electron affinity of the icosahedral C60 fullerene are calculated in the static fluctuation approximation. A graphical representation of the chemical potential equation is first obtained. The correlation function, which describes the transitions of π-electrons from one fullerene site to the nearest site, and the thermodynamic average, which characterizes the probability of detecting two π-electrons with oppositely oriented spin projections on a single fullerene site, are then calculated. The theoretically obtained values for the ionization energy of 7.57 eV and the electron affinity of 2.67 eV coincide with the experimentally observed values and demonstrate that, during photoionization or another process leading to either the acquisition or loss of a π-electron, the fullerene responds to external perturbations as a single system of strongly correlated π-electrons.

## Introduction

Electron affinity (EA) and ionization energy (IE) are fundamental characteristics of molecules [1], and therefore, their calculation is given considerable importance. Electron affinity and ionization energy play a crucial role in understanding both the type and number of bonds formed during the formation of a molecule or solid. It is equally important to understand their influence on the observed properties of various materials, from semiconductor heterostructures and devices [2, 3] to light-harvesting energetic materials [4-6] and catalysis [7]. The use of EA and IE for identifying semiconductor candidates for water splitting is also important [3].

The response of a many-particle system to an external perturbation such as photoionization is a very useful tool for studying the dynamics of correlated electrons [8], which is especially important in the case of the fullerene $C_{60}$ molecule, since the π-electron system is a strongly correlated system. It was emphasized in [8] that the presence of a large number of delocalized mobile electrons in the fullerene and the non-planar topology of the system makes the many-particle processing of electron excitations in $C_{60}$ a challenging task. According to [3], a computational method capable of determining both the EI and the EA in molecular and solid-state systems will be crucial for understanding the role of many-particle interactions in key properties of strongly correlated systems, including excited states and charge transfer effects. The aim of this work is to theoretically calculate the EI and EA in the case of the fullerene $C_{60}$. Before starting to calculate these quantities, it is necessary to understand how the EI and EA are measured experimentally.

In [9], it is noted that an important aspect that should be taken into account when studying the processes occurring in $C_{60}$ during ionization is the presence of a large number of degrees of freedom in $C_{60}$. Moreover, in fullerite, interactions between fullerene molecules are observed. To study an isolated molecule, it must be transferred to the gas phase; the material is evaporated at a furnace temperature of approximately 770 K. In [10], the furnace temperature was higher: "$C_{60}$ molecules were evaporated in a furnace with resistive heating at a temperature of approximately 873 K and ionized by monochromatic synchrotron radiation." In [11], according to the first two figures, the furnace temperature was 932 K.

Let us consider the electronic structure of the $C_{60}$ molecule. A fairly detailed review of the electronic structure of carbon nanosystems for the ten years following the discovery of buckminsterfullerene [12] is given in the book [13], and for the next 10 years – in the review article [14]. In these works, [13, 14] it was concluded that the energy levels of π-electrons are higher than the energy levels of σ-electrons, therefore π-electrons are responsible for the transport properties in fullerene. Since, when π-electrons transition from a site to an adjacent site of a fullerene, two π-electrons with oppositely oriented spin projections may end up on a fullerene site, and the Coulomb potential in this case is quite significant, it becomes necessary to take into account in the Hamiltonian, along with the transitions of π-electrons, the Coulomb repulsion of π-electrons. In 2007, an article [15] was published, devoted to the study of the structural elements of fullerene $C_{60}$ as a system of π-electrons with strong electron correlations. In works [16, 17], the energy spectrum of fullerene C60 and the optical absorption spectrum were calculated. The obtained optical absorption spectrum made it possible to explain the details of the experimental absorption spectra described in [18-

a] Prof. G.I. Mironov
Department of Physics and Materials Science
Mari State University
Kremlevskaya Street 44, Yoshkar-Ola (Russia)
 E-mail: mirgi@marsu.ru.

20], the obtained peaks in the absorption spectrum were in good agreement with the absorption peaks in [18-20].

Let us first obtain an equation for the chemical potential for the fullerene C$_{60}$; for this purpose, we will represent the Hamiltonian of the π-electron system in the form:

$$\hat{H} = \hat{H}_0 + \hat{V}, \quad (1)$$

$$\hat{H}_0 = \varepsilon \sum_{\sigma, f=1}^{60} \hat{n}_{f\sigma} + \sum_{\sigma, f \neq l} B_{fl}(a^+_{f\sigma} a_{l\sigma} + a^+_{l\sigma} a_{f\sigma}).$$

$$\hat{V} = U \sum_{f=1}^{60} \hat{n}_{f\uparrow} \hat{n}_{f\downarrow}.$$

The first term in $\hat{H}_0$ represents the self-energy of the π-electrons. This term is necessary to account for the constancy of the number of electrons in the system; we need it to derive the equation for the chemical potential. The second term describes the transitions of π-electrons from one site to an adjacent site; this is possible because the wave functions of π-electrons at adjacent sites overlap. $\hat{V}$ describes the Coulomb repulsion energy of two π-electrons with different spin projections, which, due to electron hopping, end up at the same site in the fullerene.

Having written down the equations of motion for 60 π-electron creation operators and having solved these equations in the static fluctuation approximation (the calculation method in the static fluctuation approximation is described quite well, see, for example, [15-17]), we obtain the following expression for the anticommutator Green's function:

$$\langle\langle a^+_{f\uparrow} | a_{f\uparrow} \rangle\rangle_E =$$

$$\frac{i}{2\pi} \sum_{\alpha=1}^{2} \left\{ \frac{4/120}{E - (\varepsilon_\alpha - 2{,}561B)} + \frac{3/120}{E - (\varepsilon_\alpha - 2{,}618B)} \right.$$

$$+ \frac{3/120}{E - (\varepsilon_\alpha - 0{,}382B)} + \frac{3/120}{E - (\varepsilon_\alpha + 1{,}820B)}$$

$$+ \frac{5/120}{E - (\varepsilon_\alpha - 1{,}618B)} + \frac{3/120}{E - (\varepsilon_\alpha + 2{,}757B)}$$

$$+ \frac{3/120}{E - (\varepsilon_\alpha - 1{,}438B)} + \frac{9/120}{E - (\varepsilon_\alpha + B)}$$

$$+ \frac{4/120}{E - (\varepsilon_\alpha + 1{,}562B)} + \frac{5/120}{E - (\varepsilon_\alpha - 1{,}303B)}$$

$$+ \frac{4/120}{E - (\varepsilon_\alpha - 2B)} + \frac{5/120}{E - (\varepsilon_\alpha + 0{,}618B)}$$

$$+ \frac{5/120}{E - (\varepsilon_\alpha + 2{,}303B)} + \frac{1/120}{E - (\varepsilon_\alpha + 3B)}$$

$$\left. + \frac{3/120}{E - (\varepsilon_\alpha - 0{,}139B)} \right\}. \quad (2)$$

In equation (2):
$$\varepsilon_\alpha = \begin{cases} \varepsilon, & \alpha = 1 \\ \varepsilon + U, & \alpha = 2 \end{cases}.$$

The poles of Green's function (2) define the energy spectrum of the C$_{60}$ fullerene, and the numerators of the fractions characterize the probability of finding a π-electron at the corresponding energy level. For example, the penultimate fraction in (5) indicates that an electron with spin projection ↑ can be found at the $\varepsilon_\alpha + 3B$ energy level with a probability of 1/120.

Using the fluctuation-dissipation theorem, we can pass from the Fourier transform of the anticommutator Green's function (2) to the thermodynamic average $\langle \hat{a}^+_{f\uparrow} \hat{a}_{f\uparrow} \rangle = \langle \hat{n}_{f\uparrow} \rangle$. Summing these average values over all fullerene sites, taking into account the two possible orientations of the spin projection, we obtain the following expression for the total number of π-electrons $N_e$:

$$N_e = \sum_{\alpha=1}^{2} \left( \frac{4}{e^{\beta(\varepsilon_\alpha - 2{,}561B)} + 1} + \frac{3}{e^{\beta(\varepsilon_\alpha - 2{,}618B)} + 1} \right.$$

$$+ \frac{3}{e^{\beta(\varepsilon_\alpha - 0{,}382B)} + 1} + \frac{3}{e^{\beta(\varepsilon_\alpha + 1{,}820B)} + 1} + \frac{5}{e^{\beta(\varepsilon_\alpha - 1{,}618B)} + 1}$$

$$+ \frac{3}{e^{\beta(\varepsilon_\alpha + 2{,}757B)} + 1} + \frac{3}{e^{\beta(\varepsilon_\alpha - 1{,}438B)} + 1} + \frac{9}{e^{\beta(\varepsilon_\alpha + B)} + 1}$$

$$+ \frac{4}{e^{\beta(\varepsilon_\alpha + 1{,}562B)} + 1} + \frac{5}{e^{\beta(\varepsilon_\alpha - 1{,}303B)} + 1} + \frac{4}{e^{\beta(\varepsilon_\alpha - 2B)} + 1}$$

$$+ \frac{5}{e^{\beta(\varepsilon_\alpha + 0{,}618B)} + 1} + \frac{5}{e^{\beta(\varepsilon_\alpha + 2{,}303B)} + 1} + \frac{1}{e^{\beta(\varepsilon_\alpha + 3B)} + 1}$$

$$\left. + \frac{3}{e^{\beta(\varepsilon_\alpha - 0{,}139B)} + 1} \right), \quad (3)$$

where $\beta = 1/kT$ is the reciprocal temperature.

Equation (3) is an equation for the chemical potential. In this formula, the integers in the numerator are the degeneracy multiplicities of the energy levels in the spectrum of the C$_{60}$ fullerene molecule (see, for example, [17]).

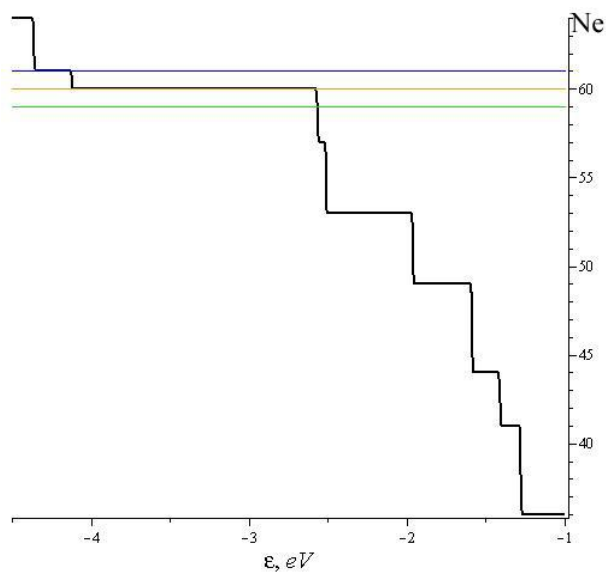

Fig. 1. Graphical representation of the equation for chemical potential.

A graphical representation of the chemical potential equation is obtained as follows: in the equality for $N_e$, we set the parameters of the Hubbard model to be $U = 7.06 \; eV, B = -0.98 \; eV$. We determined these parameters from the condition that the distance Δ between LUMO and HUMO, as follows from experiments, is equal to $1.55\ eV$. We take the reciprocal temperature $\beta = 500\ eV^{-1}$. In this case, we obtain the dependence $N_e = f(\varepsilon)$, $f(\varepsilon)$ is an analytical function. In order to understand the behavior of this function, a graph of $N_e = f(\varepsilon)$ is usually plotted. The graph of this function is shown in Fig. 1; this graph is called a graphical representation of the chemical potential equation. In this graph, the self-energy of a π-electron is plotted along the abscissa axis, and the number of π-electrons is plotted along the ordinate axis. For the subsequent calculations, we need to know three values of the self-energy. In the case of the electrically neutral fullerene $C_{60}$, there are 60 π-electrons per 60 fullerene sites. In this case, according to the graph in Fig. 1, we can choose any value of ε within the plateau for $N_e = 60$; we will take the value of $-3.53\ eV$ at the center of the plateau. The eigenenergy ε for the positively charged fullerene ion $C_{60}^+$ is uniquely determined to be $-2.57\ eV$. For the negatively charged ion $C_{60}^-$, we choose $\varepsilon = -4.35\ eV$.

Let us calculate the correlation function, for example, $\langle a_{1\uparrow}^+ a_{2\uparrow}\rangle$, characterizing the transition of a π-electron with the upward spin projection from site 2 to the neighboring site 1. To do this, we first calculate the Fourier transform of the anticommutator Green's function $\langle\langle a_{1\uparrow}^+|a_{2\uparrow}\rangle\rangle_E$ by analogy with formula (2), then, using the fluctuation-dissipation theorem, we calculate the thermodynamic average $\langle a_{1\uparrow}^+ a_{2\uparrow}\rangle$. Note that the structure of the C60 fullerene is such that the probabilities of electron transfer from any site to a neighboring site of the fullerene are the same. The thermodynamic average $\langle a_{1\uparrow}^+ a_{2\uparrow}\rangle$ will have the form:

$$\langle a_{1\uparrow}^+ a_{2\uparrow}\rangle =$$

$$= 2\sum_{\alpha=1}^{2}\left(\frac{0{,}0029\,n_\alpha}{e^{\beta(\varepsilon_\alpha+1{,}820B)}+1} - \frac{0{,}0186\,n_\alpha}{e^{\beta(\varepsilon_\alpha-1{,}618B)}+1} + \frac{0{,}0323\,n_\alpha}{e^{\beta(\varepsilon_\alpha+1{,}562B)}+1}\right.$$

$$+ \frac{0{,}0186\,n_\alpha}{e^{\beta(\varepsilon_\alpha+0{,}618B)}+1} - \frac{0{,}0226\,n_\alpha}{e^{\beta(\varepsilon_\alpha-0{,}138B)}+1} + \frac{0{,}0226\,n_\alpha}{e^{\beta(\varepsilon_\alpha+2{,}756B)}+1}$$

$$- \frac{0{,}0029\,n_\alpha}{e^{\beta(\varepsilon_\alpha-1{,}438B)}+1} + \frac{0{,}0083\,n_\alpha}{e^{\beta(\varepsilon_\alpha+3B)}+1} - \frac{0{,}0250\,n_\alpha}{e^{\beta(\varepsilon_\alpha-0{,}382B)}+1}$$

$$- \frac{0{,}0222\,n_\alpha}{e^{\beta(\varepsilon_\alpha-2B)}+1} - \frac{0{,}0324\,n_\alpha}{e^{\beta(\varepsilon_\alpha-2{,}562B)}+1} + \frac{0{,}0361\,n_\alpha}{e^{\beta(\varepsilon_\alpha+B)}+1}$$

$$\left.- \frac{0{,}0250\,n_\alpha}{e^{\beta(\varepsilon_\alpha-2{,}618B)}+1} + \frac{0{,}0293\,n_\alpha}{e^{\beta(\varepsilon_\alpha+2{,}303B)}+1} - \frac{0{,}0015\,n_\alpha}{e^{\beta(\varepsilon_\alpha-1{,}303B)}+1}\right). \quad (4)$$

In equation (4):
$$n_\alpha = \begin{cases} 1 - \langle n_{1\downarrow}\rangle, & \alpha = 1 \\ \langle n_{1\downarrow}\rangle, & \alpha = 2 \end{cases}.$$

The thermodynamic average $\langle n_{1\downarrow}\rangle$ is the average value of the number of electrons with the spin-down projection on the first site of the fullerene.

Figure 2 shows the correlation function $W = \langle a_{1\uparrow}^+ a_{2\uparrow}\rangle$ as a function of the ratio of the Coulomb potential U to the transfer integral B. In our case, the parameters of the Hubbard model are $U = 7.06\ eV$, $B = -0.98\ eV$, then the ratio of the Coulomb potential U to the transfer integral B, taking into account the minus sign in front of the ratio, is 7.2. Therefore, we will be most interested in the region when this ratio is greater than six; in this case, the regime of strong electron correlations is realized.

In the case of neutral fullerene C60, the probability of an electron jumping from site to site in this region is very low due to the fact that the Coulomb potential U significantly exceeds the transfer integral; a π-electron is practically unable to overcome the potential barrier to reach an adjacent site. The behaviour of curve 1 for the case of a positively charged $C_{60}^+$ ion is quite understandable, since an electron will always be missing at one of the fullerene sites, meaning that the potential barrier will be virtually absent on the path of a π-electron moving to this site. The behavior of curve 3, corresponding to the negatively charged $C_{60}^-$ ion, in this region can be interpreted as follows. On average, there is one π-electron at each site in this case, and there is also an "extra" sixty-first π-electron. The behavior of curve 2 shows that the transition of a π-electron from a site with one π-electron to an adjacent site with one electron is practically impossible. The remaining case is when a transition, in accordance with the Pauli exclusion principle, occurs from a site containing two π-electrons to an adjacent site with a single π-electron. In this case, the π-electron moving to the adjacent site is accelerated by the Coulomb field of the other π-electron at that site, resulting in a significantly higher electron energy, and therefore the π-electron relatively easily overcomes the Coulomb barrier.

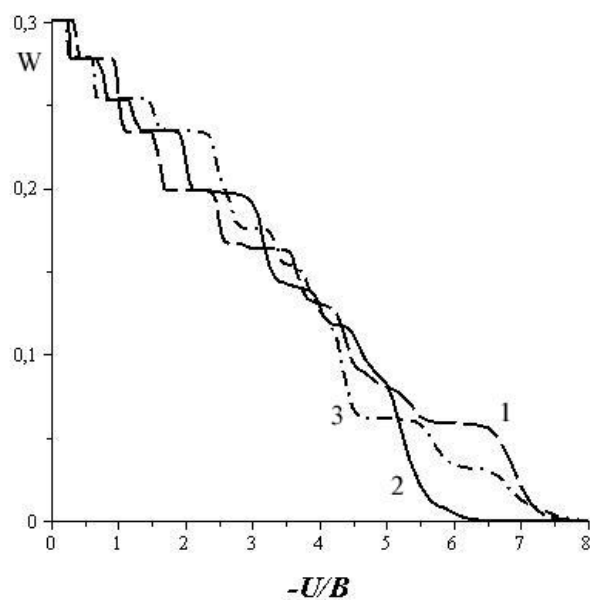

Fig. 2. Correlation function $W = \langle a_{1\uparrow}^+ a_{2\uparrow}\rangle$, describing the probability of a π-electron transition from site 2 to the neighboring site 1 of fullerene C60 depending on the ratio of the Coulomb potential U to the transfer integral B. 1 – ion $C_{60}^+$, 2 – neutral fullerene $C_{60}$, 3 – ion $C_{60}^+$.

Let us now calculate in a similar way the correlation function $\langle \hat{n}_{1\uparrow}\hat{n}_{1\downarrow}\rangle$, which characterizes the probability of finding two π-electrons at one site; it has the form:

$$\langle \hat{n}_{1\uparrow}\hat{n}_{1\downarrow}\rangle = 2\left(\frac{0{.}0250\,\langle \hat{n}_{1\downarrow}\rangle}{e^{\beta(\varepsilon+U+1{,}820B)}+1} + \frac{0{.}0250\,\langle \hat{n}_{1\downarrow}\rangle}{e^{\beta(\varepsilon+U+2{,}757B)}+1}\right.$$

$$+ \frac{0{.}0250\,\langle \hat{n}_{1\downarrow}\rangle}{e^{\beta(\varepsilon+U-0{,}139B)}+1} + \frac{0{.}0250\,\langle \hat{n}_{1\downarrow}\rangle}{e^{\beta(\varepsilon+U-1{,}438B)}+1} + \frac{0{.}0750\,\langle \hat{n}_{1\downarrow}\rangle}{e^{\beta(\varepsilon+U+1B)}+1}$$

$$+ \frac{0{.}0333\,\langle \hat{n}_{1\downarrow}\rangle}{e^{\beta(\varepsilon+U-2B)}+1} + \frac{0{.}0083\,\langle \hat{n}_{1\downarrow}\rangle}{e^{\beta(\varepsilon+U+3B)}+1} + \frac{0{.}0333\,\langle \hat{n}_{1\downarrow}\rangle}{e^{\beta(\varepsilon+U+1{,}562B)}+1}$$

$$+ \frac{0.0333 \langle \hat{n}_{1\downarrow} \rangle}{e^{\beta(\varepsilon+U-2{,}562B)}+1} + \frac{0.0417 \langle \hat{n}_{1\downarrow} \rangle}{e^{\beta(\varepsilon+U+2{,}303B)}+1} + \frac{0.0417 \langle \hat{n}_{1\downarrow} \rangle}{e^{\beta(\varepsilon+U-1{,}303B)}+1}$$

$$+ \frac{0.0417 \langle \hat{n}_{1\downarrow} \rangle}{e^{\beta(\varepsilon+U+0{,}618B)}+1} + \frac{0.0417 \langle \hat{n}_{1\downarrow} \rangle}{e^{\beta(\varepsilon+U-1{,}618B)}+1} + \frac{0.0250 \langle \hat{n}_{1\downarrow} \rangle}{e^{\beta(\varepsilon+U-0{,}382B)}+1}$$

$$+ \frac{0.0250 \langle \hat{n}_{1\downarrow} \rangle}{e^{\beta(\varepsilon+U-2{,}618B)}+1} \bigg).$$

The dependence of the correlation function $P = \langle \hat{n}_{1\uparrow} \hat{n}_{1\downarrow} \rangle$ shown in Fig. 3 on the ratio of the Coulomb potential to the transfer integral in the region of strong correlations is quite logical – the more electrons the $C_{60}$ fullerene contains, the more likely it is that two π-electrons can be found on one site of the fullerene.

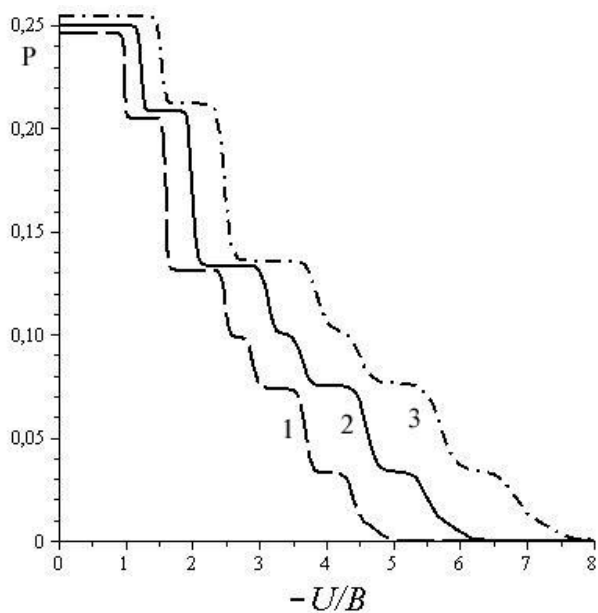

Fig. 3. Correlation function $P = \langle \hat{n}_{1\uparrow} \hat{n}_{1\downarrow} \rangle$, characterizing the probability of finding two π-electrons with opposite values of spin projections on one fullerene site depending on the ratio of the Coulomb potential U to the transfer integral B. 1 – ion $C_{60}^+$, 2 – neutral fullerene $C_{60}$, 3 – ion $C_{60}^+$.

Let us now define the energy of the ground state as the average value of the energy $E_0 = \langle \hat{H} \rangle$ at temperature $T \to 0\,K$, where the Hamiltonian $\hat{H}$ has the form [21-27]:

$$\hat{H} = \sum_{\sigma, f \neq l} B_{fl}\left(a_{f\sigma}^+ a_{l\sigma} + a_{l\sigma}^+ a_{f\sigma}\right) + U \sum_{f=1}^{60} \hat{n}_{f\uparrow} \hat{n}_{f\downarrow}.$$

In [22, 24, 27], the ground state energy was calculated by going from secondarily quantized mechanics to firstly quantized mechanics; for more details, see [26]. In [28], a calculation technique was developed and the ground state energy was calculated, remaining within the framework of second quantization, in the static fluctuation approximation. In [22-27], the results of an exact calculation of both the ground state energy of the one-dimensional Hubbard model and the average energy value at finite temperatures were presented. For comparison with the existing exact results [22-27], a special case of the one-dimensional Hubbard model in the static fluctuation approximation was considered in [28]. A comparison of the results showed that in the cases of strong and weak couplings, the ground state energies in the case of the exact solution of the one-dimensional model and in the case of the static fluctuation approximation practically coincided; in the case of intermediate correlation values, there was good quantitative agreement. The difference was that the graph for $E_0$ was slightly higher than the graph obtained from the exact solution. In this region, there was a slight overestimation of the role of the Coulomb interaction. In a fullerene, the π-electron system is strongly correlated; in this case, according to [28], the calculation error within the static approximations should be very insignificant.

The energy of the ground state (average energy value) is equal to:

$$E_0 = 360\,B\,\langle a_{1\uparrow}^+ a_{2\uparrow} \rangle + 60\,U\,\langle \hat{n}_{1\uparrow} \hat{n}_{1\downarrow} \rangle.$$

The correlation functions $\langle a_{1\uparrow}^+ a_{2\uparrow} \rangle$ and $\langle \hat{n}_{1\uparrow} \hat{n}_{1\downarrow} \rangle$ are defined above. Let us first calculate the ionization energy. To do this, we determine the ground state energy of the neutral fullerene $C_{60}$ by setting in the formulas for the thermodynamic averages $\langle a_{1\uparrow}^+ a_{2\uparrow} \rangle$ and $\langle \hat{n}_{1\uparrow} \hat{n}_{1\downarrow} \rangle$ the energy eigenvalue $\varepsilon = -3.53\,eV$ according to the graphical representation of the chemical potential equation, as noted earlier, and the average value of the π-electron concentration $\langle \hat{n}_{1\downarrow} \rangle = 1/2$. Then we calculate the energy of the ground state in the case of a positively charged fullerene ion $C_{60}^+$, setting, in accordance with the equation for the chemical potential, $\varepsilon = -2.57\,eV$, and the average value of the number of electrons at the fullerene site $\langle \hat{n}_{1\downarrow} \rangle = 59/120$. Fig. 4 shows these two graphs for the energies of the ground states.

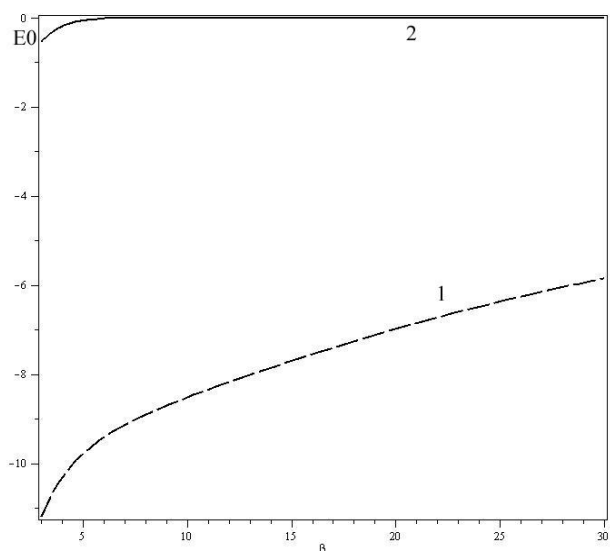

Fig. 4. The energy of the ground state in the case of a positively charged ion $C_{60}^+$ (1) and electrically neutral fullerene $C_{60}$ (2) depending on the reciprocal temperature β=1/kT.

Let us direct laser radiation at the fullerene with an energy sufficient to knock one π-electron out of the fullerene. After one electron is knocked out of the fullerene, the ground state energy of the fullerene ion with the remaining fifty-nine π-electrons will decrease. The ionization energy is the difference between the ground state energies of the electrically neutral C60 and the positively charged fullerene ion $C_{60}^+$. At the reciprocal temperature $\beta = 15.88\ eV^{-1}$ (corresponding to the temperature of the gaseous medium with fullerene molecules of $730\ K$, taking into account the slight decrease in the temperature of the gaseous medium after evaporation of the fullerene molecules from the fullerite in a furnace at a temperature of $770\ K$ [9]), the ionization energy was found to be equal to $7.57\ eV$. This value is consistent with the values of $7.57\ eV$ in the experimental work [29], $7.51\ eV$ in [30], $7.58\ eV$ in [31]. Note that in a number of works in which the ionization energy is calculated, it is defined as the difference between the energy of the positively charged $C_{60}^+$ ion and the energy of the electrically neutral fullerene, but in this case the energy is the binding energy, see, for example, formulas (1) – (3) in [32].

Electron affinity is most often determined by photoionization of a negatively charged $C_{60}^-$ fullerene ion in a gas medium using laser radiation, as a result of which the $C_{60}^-$ ion is converted into neutral C60 fullerene. After one π-electron is knocked out of the $C_{60}^-$ ion, the energy of the neutral fullerene decreases compared to the energy of $C_{60}^-$. Calculating the difference between the ground state energy of the $C_{60}^-$ ion and the ground state energy of C60, similar to calculating the ionization energy, shows that the electron affinity $E_A = 2.67\ eV$. This value is consistent with $E_A = 2.67\ eV$ in experimental studies [1, 11, 33].

Let us compare our method for calculating the energy characteristics of ionization energy and electron affinity with other methods. In [1], a simple model was proposed in which the C60 molecule was considered as a rigid conducting sphere. Formulas were obtained linking the ionization energy $E_I$ and electron affinity $E_A$ with the electron work function of a graphite monolayer $\varphi_\infty$ and the radius of the fullerene sphere $R_{60}$:

$$E_A = \varphi_\infty - \frac{1}{2R_{60}}, \quad E_I = \varphi_\infty + \frac{1}{2R_{60}}.$$

In this case, the work function of the electron from the graphite monolayer $\varphi_\infty = (E_I + E_A)/2 = 5.37\ eV$ and the electronegativity value from our calculations $(E_I + E_A)/2 = 5.12\ eV$, will differ by $0.25\ eV$. Note that the dependence of $1/R$ on the ionization energy was previously discovered in the case of small metal clusters [34-36]. In [11], when calculating the energy characteristics of fullerene, it was proposed to proceed from the fact that

$$E_A = E_{A\infty} - \frac{1}{2R_{60}}, \quad E_I = E_{I\infty} + \frac{1}{2R_{60}},$$

where $E_{A\infty}$ and $E_{I\infty}$ are the electron affinity and ionization energy for the highest fullerene with an infinite spherical radius [32]. The experimental and theoretically calculated values of $E_A$ and $E_I$ for the C60 molecule (see Fig. 4 in [11]) do not differ very much, the behavior of the fullerene in this case is close to the behavior of a simple charged sphere. However, it is clear that in the cases of electrically charged $C_{60}^-$ and $C_{60}^+$ ions, the delocalized negative or positive electric charge migrating over the fullerene sites will not be uniformly distributed over the surface of the C60 molecule at any given moment.

In [37], a model in the form of a hollow spherical conductor was associated with the fullerene; a classical analytical equation was obtained, which made it possible to estimate the electron affinity with high accuracy. But such a model did not give a satisfactory result in calculating the ionization energy, which was explained by the fact that the icosahedral fullerene is not actually a sphere; taking into account the symmetry of the fullerene leads to a splitting of the energy level at which the π-electrons with the highest energy should be located (the state with the orbital quantum number $l = 5$ is split into three states $h_u, t_{1u}, t_{2u}$ – see, for example, Fig. 3 in [17]). The consequence of such a splitting is, according to [38], that the calculation of $E_I$ will depend on the distance $\Delta = E_{60}^{LUMO} - E_{60}^{HOMO}$, where $E_{60}^{LUMO}$ and $E_{60}^{HOMO}$ are the energies of the lowest unoccupied molecular orbital and the highest occupied molecular orbital. By adding this value of the energy gap $\Delta$ to the electron affinity energy $E_A$, the value of $E_I$ was obtained, which coincided with the values of the ionization energy obtained by various methods within the framework of the density functional theory, see, for example, [39, 40]. The density functional theory shows that [37] $E_I - E_A = E_{60}^{LUMO} - E_{60}^{HOMO} + B_0$, where $B_0$ is a constant value of the order of 0.002 eV in the case of higher fullerenes. Regarding the latest calculations, we note that the system of π-electrons in the C60 molecule is a strongly correlated system [14], therefore the lower Hubbard subband in the case when there is on average one π-electron per fullerene site will be completely filled, therefore the HOMO energy level will be the energy level $t_{1u}$, as was shown in [17], and not $h_u$, as was believed in [38-40], in [17] it is shown what changes need to be made to the computing programs in order to obtain a true energy spectrum of fullerene C60, which allows us to explain the main properties of fullerene C60, including, for example, the optical absorption spectrum [16, 17]. The technique for calculating the energy spectra of π-electrons was demonstrated for the first time in [15], using the example of the structural elements of fullerene - a dimer, a pentagon and a hexagon.

Let us additionally consider computational method of an analytical nature, concerning the $E_I$ and $E_A$ in fullerene C60 [41-43]. In [41], within the framework of the Hubbard model, the energy spectrum of fullerene C60 was obtained, but for some reason, out of the sixteen energy levels of the lower Hubbard subband, only eight energy levels are considered to be included in this subband, similarly in the upper Hubbard subband. Based on the presence of these energy levels, the parameters of the system under study are calculated. In this case, along with the Coulomb repulsion energy $U$, a phenomenological fitting parameter is introduced: "$U_1$ is the energy by which $E_{60}^{HOMO}$ and $E_{60}^{LUMO}$ shift upon the removal and addition of one electron." However, the formulation of the question of the shift, for example, of $E_{60}^{HOMO}$ during the ionization of a neutral fullerene when an electron is knocked out by a quantum of the electromagnetic field, is unclear. Suppose that one electron, from the $E_{60}^{HOMO}$ level, leaves the fullerene, while other electrons still remain at the $E_{60}^{HOMO}$ level. What kind of shift in the $E_{60}^{HOMO}$ level can we be talking about? In the following work [42] the author denotes the same quantity as $U_0$ and calls it "the energy of the Coulomb interaction of an electron with charged particles entering the C$_{60}$ fullerene during ionization of a neutral molecule". A quantity such as $U_0$ can be partially discussed in the case of determining the width of the gap when measuring the differential resistance using a scanning tunneling microscope. The fact is that the observed HOMO-LUMO gap obtained by scanning tunneling microscopy is the result of the gap of the neutral C$_{60}$ fullerene plus the local Coulomb term arising from the repulsion of the tunneling electron by the electrons in the fullerene molecule. In this case, the fullerene is placed on the Au(111) surface. In this case, it is necessary to take into account the interaction of the fullerene molecule with its mirror image, obtained by "reflection" of C60 from the gold surface [44]. The broadening of the energy gap can be additionally regulated by a polarizable environment. If in the case of isolated fullerene on the Au(100) surface the LUMO resonance was detected at $\sim 1.1\ eV$ eV above the Fermi level, then the environment of C$_{60}$ by six fullerenes and triptycene molecules between the C$_{60}$ molecules leads to a shift of the LUMO resonance by approximately $0.4\ eV$ towards the Fermi level (see Fig. 11 in [44]). The works [41–43] deal with the explanation of experiments on measuring the ionization energy and electron affinity of fullerene C60 in a gaseous medium and the optical spectrum of fullerene in a gaseous medium [42] and in a solution of n-hexane [43]. In [42], the parameters of the Hubbard model and the fitting parameter $U_0$ were determined from the experimental values of $E_I$ and $E_A$ in fullerene C60, as well as using the experimental values of the peaks of three intense bands in the optical absorption spectrum of C$_{60}$. Based on the results of calculations in [42], it was concluded that the system of sixty π-electrons in a fullerene is a strongly correlated system. In [43], the same fullerene, which is a strongly correlated system, is studied as an ordinary Fermi system within the framework of the mean field in the Hubbard model. It was found that in the lower Hubbard subband, the eight lowest energy levels form a valence band filled with electrons, and the higher levels of this subband form a conduction band devoid of electrons in the ground state. Using again the experimental values of the ionization energy, the affinity energy, and the values of the three peaks of the optical absorption bands, the same as in the previous work, in [43] the parameters of the Hubbard model are calculated. After this, the optical absorption spectrum is calculated. With such results, it is impossible to speak of the significance of the works [41-43]. It should also be noted that in works [41-43] the electron self-energy ε is considered to be an fitting parameter (the average value $\varepsilon = -7.82\ eV$ with the average value of the Coulomb potential $U = 5.66\ eV$), although according to classical works [21, 23] the self-energy is determined by the value of the Coulomb potential U from the equation for the chemical potential, and in the case when there is on average one electron per site of the system, $\varepsilon = -U/2$, for example, in accordance with the graph in Fig. 1 of this work. Therefore, the problem of analytical calculation of such energy characteristics as ionization energy and electron affinity in the case of fullerene C60 was a relevant to date. In our work, a scheme for calculating these energy characteristics is proposed; using this scheme, it is possible to calculate the $E_I$ and $E_A$ in the case of any fullerenes; therefore, it is of interest to further calculate these energy characteristics in the case of the most common fullerenes and compare the calculated characteristics with already known experimental data. It is also of interest, in the context of the discussion of the influence of a tunneling electron in the case of scanning tunneling microscopy on the energy spectrum of a fullerene, to understand how the energy gap, for example, of biphenylene [45, 46], obtained from experiments on measuring differential conductivity, is related to the spectrum of elementary excitations and to the density of state of π-electrons.

## Conclusions

Thus, within the Hubbard model, in the static fluctuation approximation, it is possible to calculate both the ground-state energy and energy characteristics such as the ionization energy and electron affinity for the C$_{60}$ fullerene, with the calculated values coinciding with experimental values. The ionization energy and electron affinity are calculated as the collective response of the fullerene's system of strongly correlated π-electrons as a whole to the loss of a π-electron through photoionization or another process leading to either the acquisition or loss of a π-electron.

## Notes and references